\documentclass[reqno,oneside]{amsart}
\usepackage{amssymb}
\usepackage{amsrefs}
\usepackage[colorlinks,pagebackref,pdftitle={Hamilton--Ostrohrads'kyj relativistic spherical top dynamics},pdfdisplaydoctitle=true]{hyperref}
\makeatletter
\def\ps@pageA{
  \def\@oddhead{APPENDIX A\hss}
  \def\@oddfoot{}
  }
\def\ps@pageB{
  \def\@oddhead{APPENDIX B\hss}
  \def\@oddfoot{}
  }
\def\print@backrefs#1{%
 \space\SentenceSpace\csname br@#1\endcsname
}
\makeatother

\def\pp#1#2{\dfrac{\partial#1}{\partial#2}}
\def\bp{\boldsymbol\partial}
\renewcommand\ss[1]{{\scriptscriptstyle #1}}
\def\bd{\boldsymbol\cdot}
\def\.{\boldsymbol.\,}
\catcode`\'=13\def'{^{\boldsymbol\prime}}\catcode39=12
\catcode`\"=13\def"{{\catcode39=12'}}
\def\v{{\mathsf v}}
\def\V{\boldsymbol{\mathsf v}}
\def\p{{\mathsf p}}
\def\P{\boldsymbol{\mathsf p}}
\def\s{{\frak s}_{\ss0}}
\def\S{\boldsymbol{\mathsf s}}
\def\x{{\mathsf x}}
\def\2{^{\ss{\boldsymbol2}}}
\def\Lviv{L\hbox{\kern-1.5pt"}viv}
\def\E{\hbox{(pseudo-)} \!Euclidean }
\title[Hamilton--Ostrohrads\raise2pt\hbox{"}kyj relativistic
               spherical top dynamics]
{Hamilton--Ostrohrads\raise2pt\hbox{"}kyj approach to relativistic free
               spherical top dynamics}
\author{R.~Ya.~Matsyuk}
\address{15 Dudayev Str. 290005 \Lviv, Ukraine}
\email{romko.b.v@gmail.com, matsyuk@lms.lviv.ua}
\urladdr{\url{http://www.iapmm.lviv.ua/12/eng/files/st_files/matsyuk.htm}}
\keywords{Higher-order mechanics, Generalized Hamilton equations,
Inverse variational problem, Invariance, Relativistic top, Classical spin}
\subjclass[2010]{83C10, 70H40, 70H05, 70H50, 83A05, 49N45}
\thanks{This paper is in final form and no version of it will be submitted for
publication elsewhere}
\begin{document}
\vglue-10.2mm
{\small
\leftline{\href{http://www.univie.ac.at/EMIS/proceedings/7ICDGA/V/}{\bf DIFFERENTIAL GEOMETRY AND APPLICATIONS}}
\leftline{Proceedings of the 7\textsuperscript{th} International Conference}
\leftline{(Satellite Conference of ICM in Berlin),}
\leftline{Brno, Czech Republic, Aug.~10 -- 14, 1998.}
\leftline{Masaryk University, Brno, 1999, 547--551.}}
\vskip15mm
\begin{abstract}
Dynamics of classical spinning particle in special relativity with
Pirani constraint is a typical example of the generalized Hamilton theory
recently developed by O.~Krupkov\'a and discovers some
characteristic features of the latter.
\end{abstract}
\maketitle
\section*{Grounding}
Recent developments in Ostrohrads{"}kyj mechanics, in particular, some
substantial progress in understanding its Hamiltonian counterpart, may give
rise to the enrichment in the family of the generalized canonical dynamical
systems which describe certain processes in the real physical world. In this
report we call upon the Reader to follow the possibility of building yet
another canonical model of the free spinning particle motion in special
relativity. One way to do that is to start with the system of Dixon"s
equations \cite1 in flat space-time
$$
\advance\displaywidth by 2.7pt
\hskip140pt
\begin{cases}
\boldsymbol{\dot{\mathfrak P}}={\boldsymbol{\mathfrak 0}}
     &
     \hskip136pt
     \llap{(1a)}\\
     \vspace{1\jot}
\boldsymbol{\dot{\mathfrak S}}=2\,{\boldsymbol{\mathfrak P}}\wedge{\boldsymbol{\mathfrak u}}\,.
      &
      \hskip136pt
      \llap{(1b)}
\end{cases}
$$
The four-vector ${\boldsymbol{\mathfrak P}}$, the velocity four-vector ${\boldsymbol{\mathfrak u}}$,
and the skew-symmetric tensor ${\boldsymbol{\mathfrak S}}$ do not constitute a complete system
of variables if one wishes to put the equations \thetag{1} into the
Hamiltonian form in the usual way. On the other hand, the system \thetag{1} is
under-determined and needs to be supplemented by some constraints.

A profound classical and quantum description of the relativistic top
dynamics based upon the Dirac theory of constraints was offered by
A.~J.~Hanson and T.~Regge in \cite2, where they exploited the constraint
${\frak P}_{{\frak q}}{\frak S}^{{\frak{pq}}}={\frak0}$, sometimes referred
to as Tulczyjew supplementary condition. At the same time, some
relativistic centre-of-mass considerations, concerning the dipole model
of massive spinning particle in relativity (see \cite3 and references
therein), bring about the alternative supplementary condition,
\begin{equation}
{\frak u}_{{\frak q}}{\frak S}^{{\frak{pq}}}={\frak 0}\,, \tag{2}
\end{equation}
sometimes named Pirani supplementary condition. In the present report
I shall make an attempt to `hamiltonize" the ideology of the Pirani
constraint in contrast to what was already done with respect to the
Tulczyjew one.
\section{A brief overview of classical spinning particle settlement}
In the presence of the supplementary condition \thetag{2} it is possible to
re-solve with respect to ${\boldsymbol{\mathfrak S}}$ the following definition of the
spin four-vector ${\boldsymbol{\mathfrak s}}$,
$$
{\frak s}_{{\frak p}}=\dfrac1{2\|{\boldsymbol{\mathfrak u}}\|}\varepsilon_{{\frak{mnqp}}}
{\frak u}_{{\frak m}}{\frak S}^{{\frak{nq}}}\,,
$$
and in this way the system of equations \thetag{1b,2} may be replaced by the
following one:
$$
\advance\displaywidth by 3pt
\hskip114pt
\begin{cases}
{\boldsymbol{\mathfrak P}}=\mu_{\ss0}\,\dfrac{{\boldsymbol{\mathfrak u}}}{\|{\boldsymbol{\mathfrak u}}\|}
+\dfrac{\ast\, {\boldsymbol{\dot{\mathfrak u}}}\wedge{\boldsymbol{\mathfrak u}}\wedge
{\boldsymbol{\mathfrak s}}}{\|{\boldsymbol{\mathfrak u}}\|^{3}}
     &
     \hskip111pt
     \llap{(3a)}\\
     \vspace{1\jot}
{\boldsymbol{\dot{\mathfrak s}}}\wedge{\boldsymbol{\mathfrak u}}={\boldsymbol{\mathfrak 0}}
     &
     \hskip111pt
     \llap{(3b)}\\
     \vspace{1\jot}
{\boldsymbol{\mathfrak s}}\bd{\boldsymbol{\mathfrak u}}={\boldsymbol{\mathfrak 0}}\,.
     &
     \hskip111pt
     \llap{(3c)}
\end{cases}
$$
The quantity
$\mu_{\ss0}=\frac{\strut{\boldsymbol{\mathfrak P}}\bd{\boldsymbol{\mathfrak u}}}{\strut\|{\boldsymbol{\mathfrak u}}\|}$
entering in the expression \thetag{3a} may immediately be shown to
constitute an
integral of motion (even if we replace the right-hand side of \thetag{1a} by
some force ${\boldsymbol{\mathfrak F}}$, provided only that the condition
${\boldsymbol{\mathfrak F}}\bd{\boldsymbol{\mathfrak u}}={\boldsymbol{\mathfrak 0}}$ is obeyed). The equation \thetag{3b} may
also be given an equivalent form of
\begin{equation}
{\boldsymbol{\dot{\mathfrak s}}}
=\dfrac{{\boldsymbol{\dot{\mathfrak s}}}\bd{\boldsymbol{\mathfrak u}}}{\|{\boldsymbol{\mathfrak u}}\|^{2}}\,
{\boldsymbol{\mathfrak u}}\,, \tag{4}
\end{equation}
by means of which we deduce from \thetag{3a} that the value of the
contraction
$
{\boldsymbol{\dot{\mathfrak P}}}\bd{\boldsymbol{\mathfrak s}}\equiv({\boldsymbol{\mathfrak P}}\bd{\boldsymbol{\mathfrak s}})
{\,}{\boldsymbol{\dot{}}}
-{\boldsymbol{\mathfrak P}}\bd{\boldsymbol{\dot{\mathfrak s}}}
$
in fact equals
$\frac{\strut\mu_{\ss0}}{\strut\|{\boldsymbol{\mathfrak u}}\|}\,{\boldsymbol{\dot{\mathfrak s}}\bd{\boldsymbol{\mathfrak u}}}$,
and thus, again by means of \thetag{4}, the spin four-vector ${\boldsymbol{\mathfrak s}}$
is constant everywhere where ${\boldsymbol{\dot{\mathfrak P}}}\bd{\boldsymbol{\mathfrak s}}$ is null;
hence in the flat space-time there is no precession due to \thetag{1a}, i.e.
\begin{equation}
{\boldsymbol{\dot{\mathfrak s}}}={\boldsymbol{\mathfrak 0}}\,. \tag{5}
\end{equation}

The third-order equation of motion, obtained by substituting \thetag{3a}
into \thetag{1a}, coincides, within the realm of the Pirani supplementary
condition, with the equation suggested by Mathisson \cite4 in terms of
${\boldsymbol{\mathfrak S}}$.

Now let us fix the parametrization of the world line of the particle by means
of choosing the coordinate time as the parameter along the trajectory. We
introduce the space vs. time splitting of the variables with the help of the
following notations:
$$
     {\boldsymbol{\mathfrak u}}=(1,\V);\quad {\boldsymbol{\mathfrak P}}=({\frak P}_{\ss0},\boldsymbol{\mathsf P});\quad
     {\boldsymbol{\mathfrak s}}=(\s,\S)\,,
$$
by which the formulae \thetag{3a} and \thetag{3c} take the shape
(please notice $\V\2=\V\bd\V=\v_{a}\v^{a}=-\sum_{a=\ss1}^{\ss3}\v_{a}\v_{a}$,
although all constructions bear the same appearance independent of the
signature of the \E metrics)
$$
\advance\displaywidth by 2.7pt
\hskip62.2pt
\begin{cases}
\boldsymbol{\mathsf P}
=\dfrac{\mu_{\ss0}}{\sqrt{1+\V\2}}\,\V+\dfrac1{(1+\V\2)^{3/2}}\,\bigl(\V'\times\S
     -\s\cdot\V'\times\V\bigr)
     &
     \hskip55.7pt
     \llap{(6)}
     \\
     \vspace{1\jot}
     \s+\S\bd\V=0\,.
     &
     \hskip55.7pt
     \llap{(7)}
\end{cases}
$$
\section{Hamiltonian dynamics of free relativistic top}
We shall follow the approach of \cite5 and describe the Hamiltonian dynamics
by means of the kernel of the Lepagean differential two-form
\begin{equation}
-dH\wedge dt+d\p_{a}\wedge d\x^{a}+d\p^{\prime}_{a}\wedge d\v^{a}\tag{8}
\end{equation}
with the Hamilton function
\begin{equation}
H=\P\,\.\V - \frac{M_{\ss0}\sqrt{1+\V\2}}{\bigl(\s{}^{\ss2}+\S\2\bigr)^{3/2}}\,.\tag{9}
\end{equation}

One observation consists in  that it is possible to define such functions
$\P$ and $\P'$ of the variables $\V$ and $\V'$, that in \thetag{8} all
one-contact terms of the second order cancel out, and the expression
\thetag{8} becomes
$$
-\pp{\p_{a}}{\v^{b}}\,\omega^{a}\wedge\omega^{\prime b}
-\pp{\p^{\prime}{}_{a}}{\v^{b}}\,\omega^{\prime a}\wedge\omega^{\prime b}
-\pp{\p_{a}}{\v^{b}}\v^{\prime b}\,\omega^{a}\wedge dt
-\pp{\p_{a}}{\v^{\prime b}}\,\omega^{a}\wedge d\v^{\prime b}\,,
$$
with $\boldsymbol\omega$ and $\boldsymbol\omega'$ denoting the contact forms
of the first and of the second order resp. The functions $\P$ and $\P'$
constitute the generalized Legendre transformation, and the Lagrangian
counterpart of dynamics is described by the  Euler-Poisson expression
\begin{equation}
-\dfrac
d{dt}\,\P\Doteq-\bigl(\V'\!\.\bp_{_{\pmb{\v}}}+\V'{}'\!\.\bp_{_{\pmb{\v}'}}\bigr)\,\P\,.
\tag{10}
\end{equation}

We can suggest the following expression of the Legendre transformation
which I believe points at the adequate way to hamiltonize the
dynamics governed by the system of equations \thetag{1a\,\&\,3a},
$$
\advance\displaywidth by 2.7pt
\hskip50pt
\begin{cases}
\P=\dfrac{M_{\ss0}}{(\s{}^{\ss2}+\S\2)^{3/2}}\,\dfrac\V{\sqrt{1+\V\2}}
+\dfrac{\V'\times(\S-\s\V)}{\bigl[(\S-\s\V)\2+(\S\times\V)\2\bigr]^{3/2}}
     &
     \hskip46pt
     \llap{(11a)}\\
     \vspace{1\jot}
\P'=\dfrac{{\boldsymbol\xi}\times(\S-\s\V)}{3\,\bigl(\s{}^{\ss2}+\S\2\bigr)
\bigl[(\S-\s\V)\2+(\S\times\V)\2\bigr]^{1/2}}_{\;,}
     &
     \hskip46pt
     \llap{(11b)}
\end{cases}
$$
where
\begin{equation}
\xi_{a}
=\dfrac1{\s}\,\dfrac{(\s+\S\bd\V)\,{\mathsf s}_{a}-(\s{}^{\ss2}+\S\2)\,\v_{a}}{
(\S-\s\V)\2-({\mathsf s}_{a}-\s\v_{a})\2+(\S\times\V)\2}\,,\tag{12}
\end{equation}
and $M_{\ss0}$ is some constant number.

Looking closer at the expression \thetag{10} with $\P$ given by \thetag{11a}
convinces that the Lagrange system, defined by \thetag{10}, carries along the
primary semispray constraint (if stick to the terminology of \cite5)
\begin{equation}
\dfrac{M_{\ss0}}{\s-\S\2}\,\left[\dfrac{\S\bd\V'}{\sqrt{1+\V\2}}
-\dfrac{(\s+\S\bd\V)(\V\bd\V')}{(1+\V\2)^{3/2}}\right]=0\,.\tag{13}
\end{equation}
The expression included within square brackets in \thetag{13} presents an exact total
derivative, so we obtain the first integral of motion,
\begin{equation}
\dfrac{\s+\S\bd\V}{\sqrt{1+\V\2}}\,,\tag{14}
\end{equation}
that clearly generalizes the genuine constraint \thetag{7} which in
turn---we recall---is nothing else but the rudiment of the Pirani supplementary
condition \thetag{2}.

One would have to prove that the Hamiltonian dynamics defined by
\thetag{8,9,11} really has some connection with the classical spinning
particle dynamics given by \thetag{1a,3a,3c, and 5}. This connection clears
up in two steps. First, prove the following algebraic identity:
\begin{equation}
\dfrac{(\S-\s\V)\2+(\S\times\V)\2}{(\s{}^{\ss2}+\S\2)(1+\V\2)}
\equiv1-\dfrac{(\s+\S\bd\V)\2}{(\s{}^{\ss2}+\S\2)(1+\V\2)}\,.\tag{15}
\end{equation}
Then, multiply \thetag{11a} by the constant of motion
$
\left[\dfrac{(\S-\s\V)\2+(\S\times\V)\2}{1+\V\2}\right]^{3/2}
$
and compare with \thetag{6} to conclude that there must exist a link-up
between the constants $\mu_{\ss0}$ and $M_{\ss0}$:
\begin{equation}
\mu_{\ss0}=M_{\ss0}\,\left[1-\dfrac{(\s+\S\bd\V)\2}{(\s{}^{\ss2}
+\S\2)(1+\V\2)}\right]^{3/2}\tag{16}
\end{equation}

We may summarize the results of the preceding calculations in a couple of
statements:
\begin{itemize}
\item As far as Pirani supplementary condition is recognized, the phase space
of the free classical spinning particle may be augmented in the way that the
dynamics allows a generalized Hamiltonian description with the Hamilton
function
\begin{equation}
H=-\dfrac{M_{\ss0}}{(\s{}^{\ss2}+\S\2)^{3/2}\sqrt{1+\V\2}}\,
-\dfrac{\boldsymbol[\V',\V,\S\boldsymbol]}{\bigl[(\S-\s\V)\2+(\S\times\V)\2\bigr]^{3/2}}
\,;\tag{17}
\end{equation}
\item The mass $\mu_{\ss0}$ of the `hamiltonized' particle depends upon its
spin according to the expression \thetag{16}; it is worthwhile to
mention at this place that the Hamiltonian description of \cite2 demanded an
arbitrary dependence of the particle"s mass on its spin;

\item Any dynamical subsystem, obtained by prescribing a fixed value
to the integral of motion \thetag{14}, never is Hamiltonian by itself; in
particular, we could not have obtained a variational description of the
spinning particle motion if the constant of motion \thetag{14} had been
frozen by means of the equation \thetag{7} or, equivalently, by the demand
that $\mu_{\ss0}$ and $M_{\ss0}$ take the same value in \thetag{16};

\item The Legendre transformation, given by \thetag{11}, is not globally
defined in an intrinsic sense, as may be seen from \thetag{12}; nevertheless,
the Hamilton function is defined quite nicely via the expression \thetag{17}.
\end{itemize}

Guessing the form of the Legendre transformation \thetag{11} is equivalent to
solving the Poincar\'e-invariant inverse problem of calculus of variations
in order 3. That was treated in \cite6 and the corresponding Euler-Poisson
expression \thetag{10} found. But I did not know the appropriate expression
for the Legendre transformation until 1995 when a set of Lagrange functions
corresponding to \thetag{10} was discovered \cite7.

\clearpage
\hoffset-3truecm
\voffset-2truecm
\hsize180truemm
\vsize230truemm
\pdfbookmark{APPENDIX A. Spin dependence of test particle mass}{pgA}
\thispagestyle{pageA}
{\tt
\rightline{14$^{\rm th}$ International Conference on General Relativity and Gravitation,}
\rightline{Florence, Italy, August 6-12 1995.}
\rightline{Abstracts of Contributed Papers, p.~A.120.}
}
\medskip
\let\cal=\mathcal
\font\rmsmall=cmr8
\font\blarge=cmbx12 scaled\magstephalf
{\parindent0cm
{
\openup1\jot
\blarge\obeylines Spin Dependence of Classical Test Particle Mass in the Third-Order
Relativistic Mechanics
}

\smallskip
{\bf Roman Ya. MATSYUK}

{
\smallskip
\leftskip1.1cm
\openup-1\jot
\rmsmall\obeylines  Institute for Applied Problems in Mechanics \& Mathematics, 3-B Naukova St. 290601 \Lviv, Ukraine.
E-mail: matsyuk@lms.ipam.lviv.ua
\smallskip}}

From time to time there arise some higher-order equations of
motion in
the classical mechanics of particles. Perhaps the best known one is the
Lorentz-Dirac equation of the radiating electron. Unfortunately,
in the real physical dimension of four one may never derive a third-order
equation of motion with Lorentz symmetry from a variational
principle.\footnote{\hsize180truemm R.~Ya.~Matsyuk.
\sl Poincar\'{e}-invariant equations of motion in
lagrangian mechanics with higher derivatives.
\rm Thesis, \it Institute for Applied Problems in Mechanics and
Mathematics, Academy of Science, Ukraine. \rm \Lviv, 1984 (in Russian).}

The Lorentz-Dirac equation, however, may be put down into a more general
framing of the Dixon system of equations
$$
\hskip2cm\textstyle{1\over2}\dot S=p\wedge u\,;\quad \dot p={\cal F}
\leqno\rm(A1)
$$
by introducing the momentum
\thinspace$p={m_0u\over|u|}+{\dot S\ldotp u\over u^{\bf2}}$ \thinspace and
the spin tensor
\thinspace${1\over2}S={4\over 9}{e^4\over m_0}{u\wedge\dot u\over|u|^3}$.
\thinspace
Conversely\footnote{\hsize180truemm R.~Ya.~Matsyuk.
\it In:\/ \rm Abstracts of Contributed Papers. 11$^{\rm th}$ International
Conference on General Relativity and Gravitation. Stockholm, July~6--12,~1986.
Vol.~\uppercase\expandafter{\romannumeral2}, p.~648},
it may be cut off from~(A1) by means of the constraint
\thinspace$\dot S\ldotp u+{2e^2\over 3}\left(\dot u
-{(u\cdot\dot u)\over u^{\bf2}}\,u\right)=0$ \thinspace
and defining the force in~(A1) by
\thinspace${\cal F}=-{\scriptstyle2\over\scriptstyle3}\,e^2\kappa^2u$,
\thinspace
where $\kappa$ stands for the curvature of the particle's world line.

These considerations suggest that there may still exist a third-order
equation of motion with the Lorentz symmetry, derivable from a variational
principle if only some spin variable were retained in it.
In course of developing some previous work\footnote{\hsize180truemm R.~Ya.~Matsyuk.
Dokl. Akad. Nauk SSSR, {\bf285} (1985), No.~2, 327--330, English translation: Soviet Phys. Dokl., {\bf30} {1985}, No.~11, 923--925, MR0820861(87d:70029).}
we offer here the following form of the spinning particle equations of
motion
$$
\displaylines{
{\rm(A2)}\hskip2cm\hfill{\mu_0\over|\sigma|^3}\left[{(u\cdot\dot u)\over|u|^3}u-
{\dot u\over|u|}\right]
-{\ast\,\ddot u\wedge u\wedge\sigma\over|\sigma\wedge u|^3}
+3\,{\ast\,\dot u\wedge u\wedge\sigma\over|\sigma\wedge u|^5}
\,(\sigma\wedge\dot u)\cdot(\sigma\wedge u)
={|u|\over|\sigma\wedge u|^3}{\cal F}\,,\cr
{\rm(A3)}\hskip2cm\hfill\dot\sigma\wedge u=0\hfill\cr
}$$
where we have set \thinspace${\cal F}_\alpha
={1\over2}R_{\alpha\beta}{}^{\gamma\delta}u^\beta S_{\gamma\delta}$.
\thinspace

The system of equations (A2,~A3) may be put into the Dixon form~(A1)
by introducing the spin tensor
\thinspace${1\over2}S={\ast\,u\wedge\sigma\over|u|}$ \thinspace
{\it and imposing the constraint}
$$
\hskip2cm(\sigma\cdot u)\,=\,0\,.\leqno\rm(A4)
$$
Conversely, if we introduce the spin four-vector $\sigma$, then the Pirani
auxiliary condition \thinspace$S\ldotp u\,=\,0$ \thinspace will cut off
the system of equations (A2--A4) from the Dixon system~(A1). This way the Dixon
system~(A1) may be thought of as a ``covering equation'' both to the
Lorentz-Dirac equation and to the equation~(A2). But the latter {\it admits a
lagrangian}.
In the flat Lorentz space-time the lagrangian for the equation~(A2) reads
$$
\hskip2cm L={\mu_0\over|\sigma|^3}\,|u|+{L_{(\alpha)}\over\sigma^{\bf2}|\sigma
\wedge u|}\,,
\hbox{ \ with \ }
L_{(\alpha)}\,=\,
{\ast\,\dot u\wedge u\wedge\sigma\wedge e_{(\alpha)}\over(u_{\alpha}\sigma
-\sigma_{\alpha}u)^{\bf2}-(\sigma\wedge u)^{\bf2}}
\,\bigl(\sigma^{\bf2}u_{\alpha}+\,(\sigma\cdot u)\,\sigma_\alpha\bigr)\,,
$$
where $e_{(\alpha)}$ denotes the $\alpha$-component of the Lorentz frame.
Every $L_{(\alpha)}$ will do because each differs from the others by
some total derivative.
This holds irrespective of the constraint~(A4).

From the
variational point of view it is even more natural to consider the equation~(A2)
also in the region outside the manifold defined by the constraint~(A4).
If we assume that~(A1) together with the Pirani auxiliary condition governs
the motion
of a spinning particle with the constant mass $m_0$, then it follows that
the equation~(A2) admits (at least in the flat space-time) a spectrum of
the particles with the variable mass
\thinspace$m
=\mu_0\left[1-{(\sigma\cdot u)^2\over\sigma^{\bf2}u^{\bf2}}\right]^{3\over2}
$. \thinspace
This quantity $m$ is a constant of motion if only $\dot\sigma=0$ and
${\cal F}=0$. The shortcoming of our approach is that we freeze the spin
variable in the variational procedure. We shall try to overcome this
threshold in future work. On the other hand, it is common that the spin
four-vector $\sigma$ does not change along the world line of the particle
in the flat space-time.
\clearpage
\vsize245truemm
\pdfbookmark{APPENDIX B. Lagrangian approach to spinning particle}{pgB}
\thispagestyle{pageB}
\catcode`\'=13\def'{\kern1.5pt^{\prime}}\catcode39=12
\catcode`\"=13\def"{{\catcode39=12'}}
{\tt
\rightline{11$^{\rm th}$ International Conference on
General Relativity and Gravitation,}
\rightline{Stockholm, Sweden, July 6-12, 1986.}
\rightline{Abstracts of Contributed Papers, vol.~\uppercase\expandafter{\romannumeral2}, p.~648.}}
\bigskip
\rightline{20:16}
\bigskip
\centerline{LAGRANGIAN APPROACH TO SPINNING OR RADIATING PARTICLE HIGHER-ORDER}
\centerline{EQUATIONS OF MOTION IN SPECIAL RELATIVITY}
\bigskip
\centerline{R.~Ya.~Matsyuk (USSR)}
\bigskip
We recognize it worthwhile to distinguish among the higher-order equations of motion those admitting a variational reformulation. Considering Dixon's form of Papapetrou equations either for free particle,
$$\hskip2cm p'=0,\quad S'=2\,p\wedge u\,,\leqno\rm(B1)$$
or introducing an external force by means of $p'=k{u'}^2u$, and imposing within the restriction $\Vert u\Vert =1$ a supplementary condition $S'u+ku'=0$, one regains the Lorentz-Dirac equation, $mu'=k(u^{\prime\prime}+{u'}^2u)$, in the second case with constant mass $m=pu$, reversing thus the considerations of A.~O.~Barut. The Lorentz-Dirac equation, however, cannot be reformulated in terms of a variational principle.* In the presence of another supplementary condition, $Su=0$, the equations~(B1) can be given the form
$$\hskip2cm mu'=Su^{\prime\prime}\;\hbox{(M.~Mathisson, 1937)},\quad u\wedge S'=0,\quad (u^2=1).\leqno\rm(B2)$$
In terms of the star operator~$*\,$, the spin four-vector $\displaystyle s={*\,(u\wedge S)\over2\Vert u\Vert}$, and complying with the obvious constraint $su=0$, equation~(B2) can be rewritten as
$$\hskip2cm mu'=*\,(u^{\prime\prime}\wedge u\wedge s),\quad (u^2=1,\quad s'=0).\leqno\rm(B3)$$
Renormalizing the mass by means of $$m=m_{*}\left(1-{(su)^2\over s^2u^2}\right)^{3/2},\leqno\rm(B4)$$
and taking $m_{*}$ to be an arbitrary but fixed constant, we state here that the following third-order equation of motion defining the same set of world lines as~(B3),
$$\hskip2cm m_{*}{u^2u'-(u'u)\,u\over\Vert s\Vert ^3\Vert u\Vert ^3} ={*\,(u^{\prime\prime}\wedge u\wedge s)\over\Vert s\wedge u\Vert ^3} -3 *\,(u'\wedge u\wedge s)\thinspace
{(s\wedge u')\cdot(s\wedge u)\over\Vert s\wedge u\Vert ^5},$$
is the Euler-Poisson equation of a parameter-invariant variational
problem. Proceeding further in eliminating the spin variables from~(B3) results in the fourth-order equation
$$\hskip2cm u^{\prime\prime\prime}={p^2\over s^2}\,u',\quad (u^2 =1),$$
which under the assumption $p\kern1.5pt^2>0$ was suggested from the higher-order Lagrangian point of view by F.~Riewe without indicating any direct relationship with Papapetrou equations.

\bigskip
REFERENCES. A.~O.~Barut. {\it In:} ``Quantum Opt. Exp. Gravity, and Meas. Theory'',
Proc. NATO Adv. Study Inst., Bad Windsheim, 1981, N.~York--London, 1983, 155--168. M.~Mathisson. Acta Phys.~Polon., {\bf6} (1937), fasc.~3, 163--200. *R.~Ya.~Matsyuk.
Dokl. Akad. Nauk SSSR, {\bf282} (1985), No.~4, 841--844, English translation: Soviet Phys. Dokl., {\bf30} (1985), No.~6, 458–460, MR0802859 (87d:70028). T F.Riewe. Il Nuovo Cim.~B, {\bf8}
(1972), No.~1, 271--277.	

\begin{bibdiv}
\begin{biblist}
\bib{1}{article}{
title={Dynamics of extended bodies in general
relativity~\uppercase\expandafter{\romannumeral1}.  Momentum and angular momentum},
author={W.~G.~Dixon},
journal={Proc. Royal Soc. London, Ser.~A.}
volume={314},
date={1970},
pages={499--527}
}

\bib{2}{article}{
title={The relativistic spherical top},
author={A.~J.~Hanson},
author={T.~Regge},
journal={Ann. Phys.}
volume={87},
number={2},
date={1974},
pages={498--566}
}

\bib{3}{book}{
title={The theory of relativity},
author={C.~M\o ller},
publisher={Springer},
address={Berlin--Heidelberg},
date={1972}
}

\bib{4}{article}{
title={Neue Mechanik materieller Systeme},
author={M.~Mathisson},
journal={Acta Phys. Polon.}
volume={6},
number={3},
date={1937},
pages={ 163--200}
}

\bib{5}{book}{
title={The geometry of ordinary variational equations},
author={Olga~Krupkov\'a},
series={Lecture Notes in Mathematics},
volume={1678},
publisher={Springer},
date={1997}
}

\bib{6}{thesis}{
title={Poincar\'e-invariant equations of motion in
Lagrangian mechanics with higher derivatives},
author={R.~Ya.~Matsyuk},
organization={Institute for Applied Problems in Mechanics and
Mathematics, Academy of Science of Ukraine, \Lviv}
type={Ph.D. Thesis},
date={1984}
pages={140}
language={Russian}
}

\bib{7}{article}{
title={Spin dependence of classical test particle mass
    in the third-order relativistic mechanics},
author={R.~Ya.~Matsyuk},
booktitle={Abstracts of Contributed Papers},
conference={
title={14\textsuperscript{th}~International Conference on General Relativity and Gravitation},
date={August 6--12, 1995},
address={Florence}
},
pages={A.120}
}

\end{biblist}
\end{bibdiv}
\end{document}